\documentclass{ws-ijmpc}
\usepackage[super]{cite}
\usepackage[breaklinks]{hyperref}
\hypersetup{colorlinks,urlcolor=black,citecolor=black,linkcolor=black,filecolor=black}
\usepackage{breakurl}
\begin{document}

\title{Order-disorder-order transitions and winning margins' scaling in kinetic exchange opinion model}
\author{Soumyajyoti Biswas}
\address{Department of Physics and Department of Computer Science and Engineering, SRM University - AP, Amaravati, Andhra Pradesh 522240, India}
\author{Mandhalapu Sree Annapurna, Venkatasubbaiah Jakkampudi, Dushyanth Yarlagadda, Bhargav
Thota}
\address{Department of Computer Science and Engineering, SRM University - AP, Amaravati, Andhra Pradesh 522240, India}

%\markboth{WSPC}{Instructions for typesetting manuscript}

\maketitle
\begin{abstract}
 The kinetic exchange opinion model shows a well-studied order disorder transition as the noise parameter, representing discord between interacting agents, is increased. A further increase in the noise drives the model, in low dimensions, to an extreme segregation ordering through a transition of similar nature. The scaling behavior of the winning margins have distinct features in the ordered and disordered phases that are similar to the observations noted recently in election data in various countries, explaining the qualitative differences in such scaling between tightly contested and land-slide election victories.  
\end{abstract}

\section{Introduction}
The quantification of opinion formation through exchanges of interacting individuals has been a long-standing effort from  physicists and social scientists \cite{oup,rmp,galam1,spl}. Specifically, the interest of statistical physicists has been to model this process as an emergent phenomenon with a rich set of observations that arise out of rather simpler rules of interaction among the agents holding different opinions. While opinion exchanges have nuances that are not easily captured through simple rules, certain situations can help model parts of it. For example, a popular choice has been to utilize the Ising symmetry in the opinion variable associated with an agent (see e.g., \cite{snj,deff,hk,toscani,lccc,tp_1,tp_2}). This simply means that the agents are having to make a binary choice, or be neutral, in their opinion values. It could be voting in a referendum (Brexit is an example, see e.g., \cite{brexit}), a voting with two major candidates (The U.S. elections are good approximations, see e.g., \cite{cg1,cg2}) or simply any other yes/no choice.   

In the above mentioned examples, the interacting agents can reach a state of consensus where majority of the agents prefer one choice, or a fragmented state where no clear consensus is likely. Nevertheless, a narrow margin can exist even in such cases between the two groups due to finite size fluctuations. This transition, from consensus based order to disorder with no clear consensus have been widely investigated using tools of critical phenomena. Many of the realistic properties observed in the real data from elections in various contexts were compared with such models. The interests span from details of critical scaling \cite{nuno1,nuno2,kb}, effect of topology of the underlying interaction network \cite{ba_sz,lima,solomon} to upper and lower critical dimensions of the models \cite{sudip}, candidature based on populations and spending abilities \cite{can1,can2,can3}, the scaling behavior of vote distributions among candidates \cite{vote_dist,ac}, spatial correlations and comparisons with demographic data \cite{noisy_voter} to time of switching of the global opinion between the two choices \cite{brexit} etc. Recently a universal scaling behavior of the winning margins was observed in election data from various countries \cite{san}.  

In this work we use the kinetic exchange opinion model \cite{bcs} with binary interactions and a single tunable parameter representing the nature of the interactions between the agents. Such models are in the genre of kinetic models of ideal gas atoms, applied to social dynamics of wealth exchanges \cite{cup}. Here, of course, no conservation laws apply, such as the momentum and money conservations in the ideal gas and wealth exchange models respectively. But a bound on the extreme positive and negative values (arbitrarily set to $\pm 1$) is enforced. The resulting dynamics show a spontaneous symmetry breaking transition at a critical noise strength, with exponent values in the Ising universality class \cite{sudip,kb}. It is unclear, however, if the upper critical dimension is the same as that of the Ising model \cite{sudip}. Similar questions were also addressed in the case of other models with Ising symmetry, such as the noisy voter model (see e.g., \cite{nv1,nv2}).

Here we show that a further increase beyond the order-disorder transition point in the noise parameter takes the population through complete randomness, where each agent is equally likely to be in any of the groups ($\pm 1$ and $0$) at any time, to a complete segregation between two groups of opposing polarity (only $\pm 1$, as the neutral fraction disappears), like echo-chambers \cite{echo}. This is again a transition of the same universality class that manifests itself through an anti-ferromagnetic ordering when implemented on a two dimensional square lattice. Furthermore, we divide the agents into different groups, representing electoral constituencies, and  study the scaling behavior of the wining margins in the different constituencies. A qualitative shift in the scaling function exists when computed in the ordered and in the disordered phases. These are similar to such changes observed in the real data for closely contested elections and one sided victory. 

\section{Model and methods}
The kinetic exchange opinion model consists of $N$ agents, arranged either on a regular lattice or in a fully connected manner. Each agent has an opinion value $o_i(t)$ associated with them at any time $t$, with $i=1,2,\dots, N$. A time step consists of $N$ binary interactions between two agents, following
\begin{equation}
    o_i(t+1)=o_i(t)+\mu o_j(t),
\end{equation}
where $\mu$ represents the nature of interactions and can take values $-1$ or $+1$ with probabilities $p$ and $1-p$ respectively. Opinion values are initialized randomly between $\pm 1$ and $0$, and a bound is enforced at the extreme ends i.e., after the exchange for $o_i(t+1) > 1$ it is set to $+1$ and for $o_i(t+1)<-1$ it is set to $-1$. The emergent order in the model can be measured through 
\begin{equation}
O=\frac{1}{N}|\sum\limits_{i=1}^N o_i(t\to \infty)|,
\label{op_n}
\end{equation}
showing an order (consensus) to disorder (random) transition at a critical noise strength $p_c$. The choices of the interacting pair $(i,j)$ are random for the mean field version of the model, and are randomly selected nearest neighbors when an underlying topology exists. The critical noise strength in the mean field limit can be shown to be $p_c=1/4$ \cite{bcs}, and for square lattice it is numerically seen to be around $p_c\approx 0.11$ \cite{sudip}. The critical point, of course, is sensitive to topology, but the universality class is only dependent on the dimensions. 

Additionally, we also measure the staggered opinion, defined similarly as the staggered magnetisation for the Ising anti-ferromagnets
\begin{equation}
    O_s=\frac{1}{N}|\sum\limits_{i=1}^N(-1)^{x_i+y_i}o_i(t\to\infty)|,
    \label{op_stag}
\end{equation}
where $(x_i,y_i)$ denotes the coordinates of the $i$-th agent on a square lattice. This quantity can capture a checker-board order developing in the system for high values of the noise parameter $p_{c_2}=1-p_{c_1}$, as discussed later on. 

We then define a set of groups of agents, representing different electoral constituencies. In the square lattice geometry, these are simply $M\times M$ subgroups, and in the mean field limit just groups of size $M$. These grouping has no effect on the interactions between the agents i.e., agents belonging to different groups can interact as usual. But these groups are used to measure the winning margins (difference between the number of agents having positive and negative opinion values) in each constituency. The statistical properties of these margins are then compared between different phases of the model i.e., for different values of the noise parameter $p$, and then compared with recent observations in the real data. 

All simulations done here are either on a square lattice with periodic boundary conditions or in the mean field limit. The initial conditions are $\pm 1$ opinion values for all agents, and one Monte Carlo time step comprises of $N$ binary interactions, along with the bounds enforced as needed. The relevant quantities are measured in the long time limit when a steady state is reached. 

\section{Results}
In this section, we report the simulation results for the model in two dimensions and in the mean-field (fully connected graph) limits. It has been reported elsewhere that the model in these two limits, behave like the Ising model \cite{front}. It is not clear if the universality holds in other dimensions, but that is not going to be addressed here. Here, we first look at the order-disorder-order transition in the two-dimensional case. We then go on to divide the system into groups, representing individual voting units/constituencies and study the behavior of the winning margins, as a function of the noise parameter $p$. Finally, we compare with the real data from the U.S. and the U.K. elections and compare our model's results with different electoral conditions. 

\begin{figure*}[tbh]
\includegraphics[width=11cm]{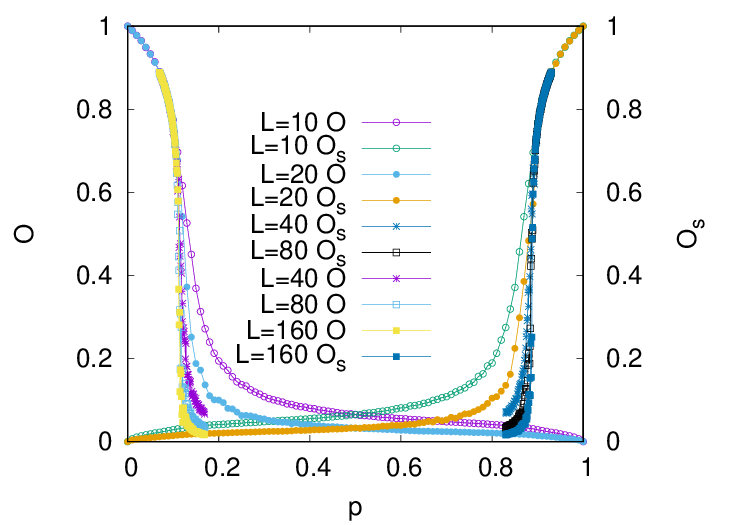}
\includegraphics[width=11cm]{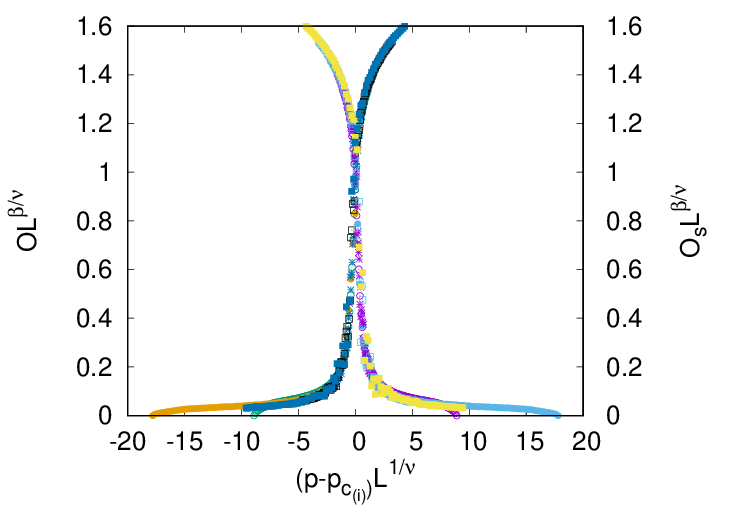}
\caption{The variations of the order parameters with the noise parameter $p$ and their finite size scaling. The top figure shows the variations of the average opinion $O$, and staggered opinion $O_s$ (defined similarly as the staggered magnetization for anti-ferromagnets). The bottom figure shows the usual finite size scaling, with $\beta=0.125$ and $\nu=1$. The two critical points are $p_{c_1}=0.12$ and $p_{c_2}=1-p_{c_1}=0.88$.}
\label{fig1}
\end{figure*}

\subsection{Order-disorder-order transition}
The consensus to non-consensus transition in this model have been studied for a long time under different topologies. Similar such transitions, driven by noise, are also reported in various other models. However, the difference between the noise parameter in this model and elsewhere is that here the noise represents discord between the interacting agents, and it is not simply randomness (see e.g., \cite{nv1,nv2}). As can be seen from the model definition, increasing the discord would prompt the interacting agents to adopt an opposing political stance, notwithstanding their earlier stance. This formulation, in its root, is the driving mechanism towards polarization rather than just disorder or randomness. Indeed, the value $p=1/2$ is the perfect random state, where the agents will choose randomly between the three possibilities $\pm 1$, $0$ (neutral) with equal probabilities. For $p>1/2$, the model goes towards a polarization order. 

A mean-field version is not sufficient to capture this polarization ordering. But when implemented on a square lattice, this gives rise to an anti-ferromagnetic order in the system. In the extreme case of $p=1$, an anti-ferromagnetic order is stable for a square lattice. There are no neutral agents in this case. Indeed, this anti-ferromagnetic order occurs at a symmetric points with respect to $p=1/2$, the complete randomness. If we denote the first order-disorder transition point by $p_{c_1}$, then this second ordering towards polarization would occur at $p_{c_2}=1-p_{c_1}$. In Fig. \ref{fig1}, the two transitions are captured by the total average opinion (similar to magnetisation) and the staggered opinion (see Eqs. (\ref{op_stag}) and (\ref{op_n})), respectively. They show a perfectly symmetric behavior about $p=1/2$. Both transitions also show the same finite size scaling behavior, with the same exponent values $\beta=0.125$ and $\nu=1$, as in the Ising universality class. This clarifies that the universality class of both transitions is the same, although the states of the system in both types of ordering are very different -- one having a consensus ($p<p_{c_1}$) and the other is absolute polarization ($p>p_{c_2}$), driven by the same noise parameter. Both transitions points show fluctuation divergence with same exponents (not shown) and expected to match the Ising universality class in all other aspects, at least for a square lattice.

\begin{figure*}[tbh]
\includegraphics[width=11cm]{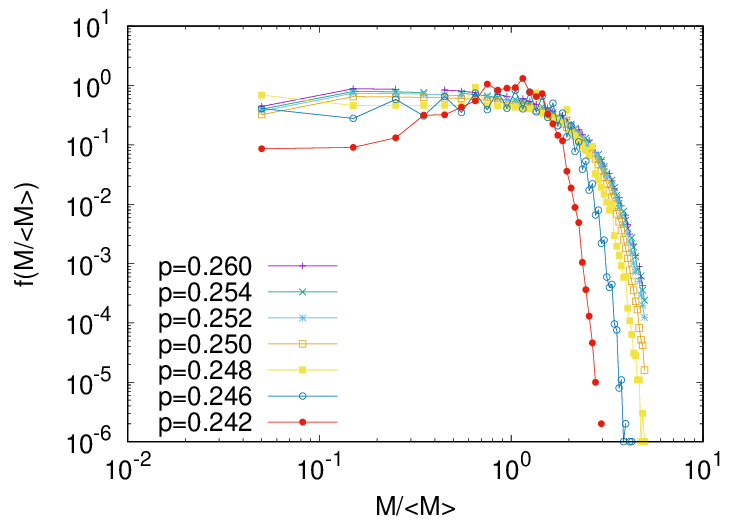}
\includegraphics[width=11cm]{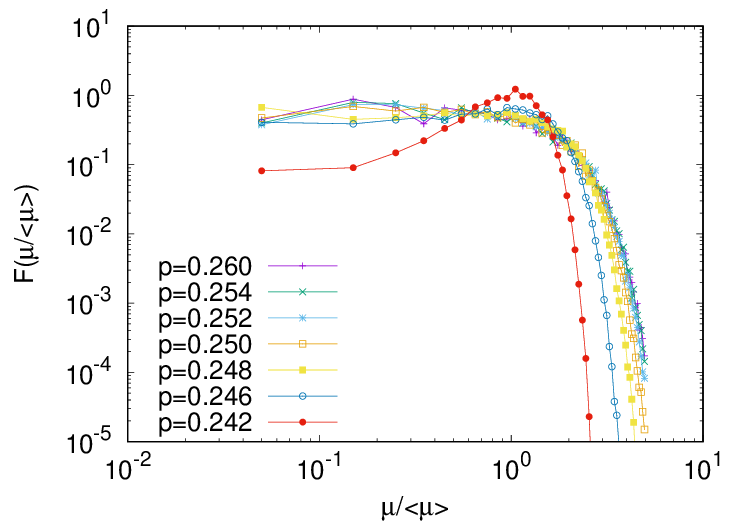}
\caption{The distribution of the winning margins $M$ and the scaled margins $\mu=M/T$ with $T=f_1+f_{-1}$ for the mean-field version of the model. The critical point is exactly known at $1/4$. The behavior of the distributions qualitatively differ on either side of the critical point.}
\label{fig2}
\end{figure*}
\begin{figure*}[tbh]
\includegraphics[width=11cm]{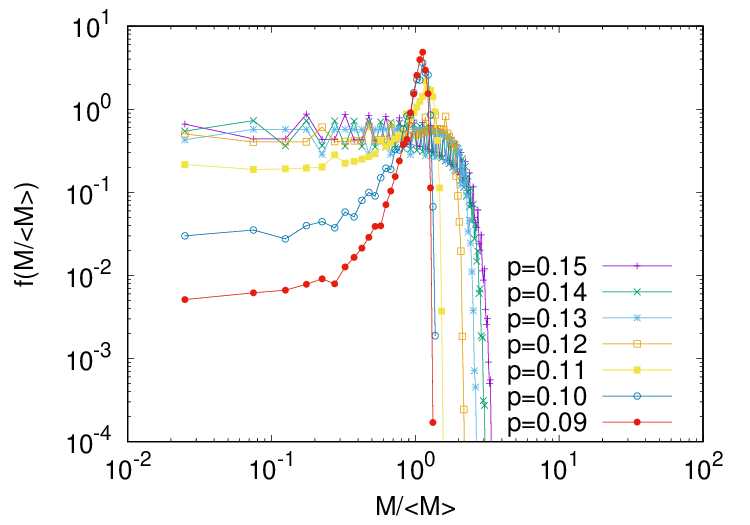}
\includegraphics[width=11cm]{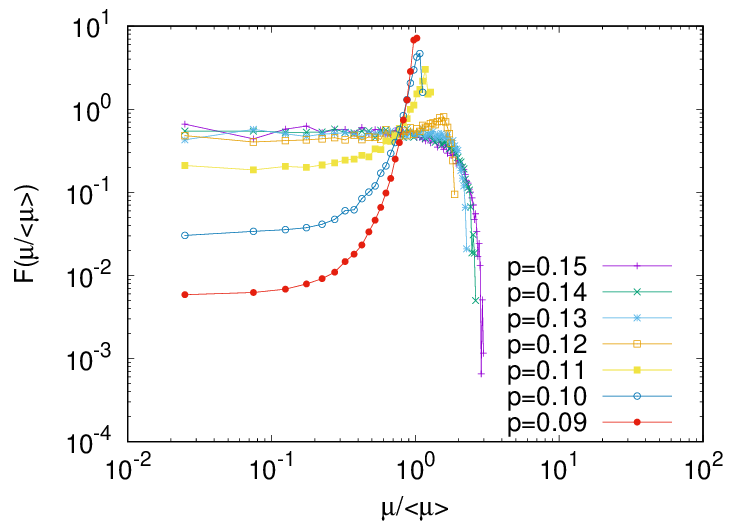}
\caption{The distribution of the winning margins $M$ and the scaled margins $\mu=M/T$ with $T=f_1+f_{-1}$ for the two-dimensional version of the model. The critical point is numerically estimated to be around $0.11$. Like in the mean-field case, the behavior of the distributions qualitatively differ on either side of the critical point.}
\label{fig3}
\end{figure*}
\begin{figure*}[tbh]
\includegraphics[width=11cm]{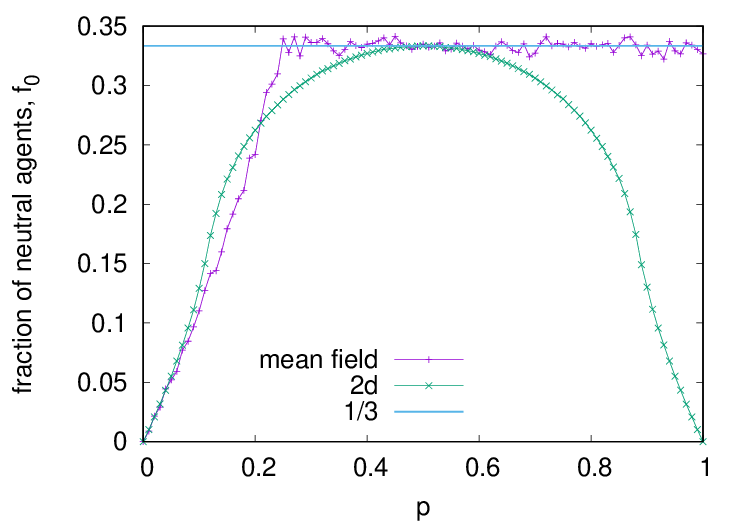}
\caption{The variation in the fraction of neutral agents are shown for the mean-field ($N=10000$) and square lattice ($N=L \times L$ with $L=100$) as a function of the noise parameter $p$. As the polarized state is formed for $p>1/2$, the neutral fraction starts decreasing for the square lattice geometry. No such ordering is possible for the mean-field case, where the fraction remains at $1/3$.}
\label{fig4}
\end{figure*}

\begin{figure}[h]
\includegraphics[width=7cm]{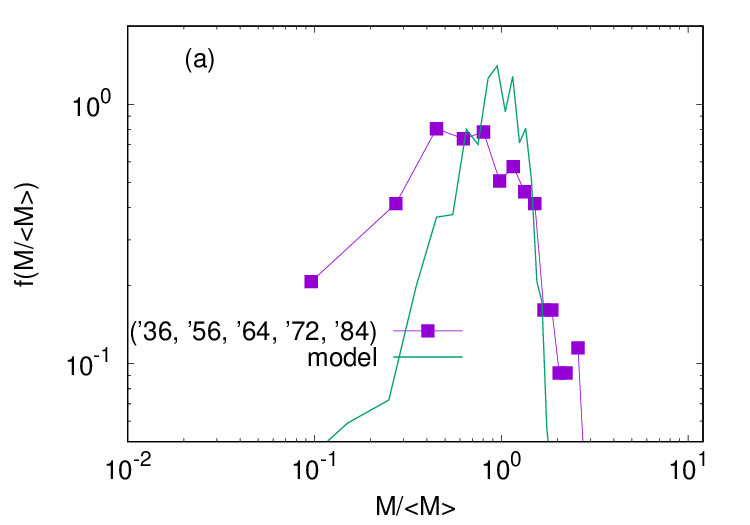}
\includegraphics[width=7cm]{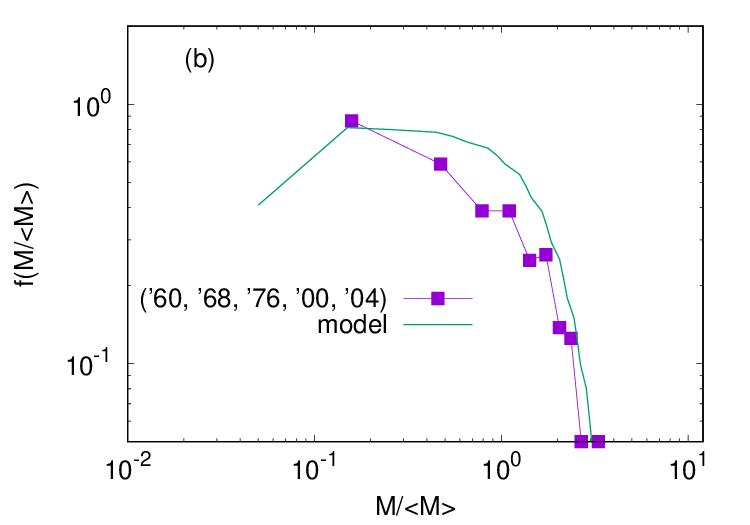}
\includegraphics[width=7cm]{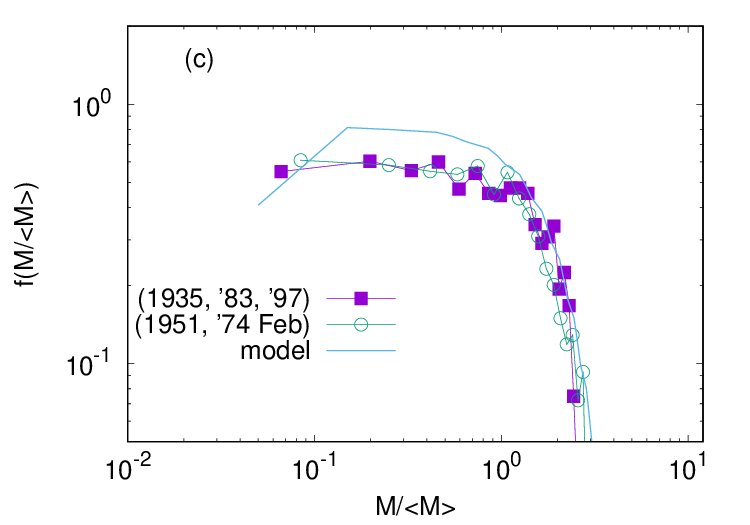}
\caption{The comparisons between the distribution of the winning margins $M$ for some of the land-slide and tightly contested elections in the U.S. and the U.K. and the same from the model simulation (in the mean field limit): (a) The distributions of margins for some of the land-slide elections in the U.S. and the same for the model for $p=0.24 (<p_c=0.25)$, (b) the same comparisons for some of the tightly fought elections in the U. S. and the model for $p=0.255 (>p_c)$, (c) plots for the U.K. elections, where the distinction is not significant and both are compared with the data from the model simulation with $p=0.255$. In all cases, there is a qualitative agreement between the real data and that from the simulations of the model.}
\label{fig5}
\end{figure}

\subsection{Winning margin distributions and voting turnout}
It has recently been shown \cite{san} that across various countries, distribution of the winning margins (or that scaled by turnout) in different constituencies show universal pattern. In some cases, however, there are significant departures, possibly associated with limited transparency of the election processes. 

It is useful, therefore, to see if this model shows such realistic distributions in terms of the winning margins. To do that, we divide the agents in the system in to sub-groups of size $K$ in the mean-field version and blocks of size $K\times K$ in the square lattice, representing different constituencies. There is no effect of this grouping in terms of their interactions, of course. However, at any time, one can calculate the fractions of agents with opinion values $\pm 1$ and $0$, represented by $f_1$, $f_{-1}$ and $f_0$. The winning margin in any constituency is given by $M=|f_1-f_{-1}|$, and turn out is $T=f_1+f_{-1}=1-f_0$, assuming that the neutral agents do not vote. Later on we broaden the interpretatation of neutral agents to include all who do not vote and also the ones who vote for any other candidate than the two dominant parties of that country (for example, voters other than Democrats and Republicans in the US elections).

In the simulations of this model in the mean-field and two-dimensions, we go on to calculate $M$, $T$ and $\mu=M/T$, and the distributions $f(M/\langle M\rangle)$ and $F(\mu/\langle\mu\rangle)$, respectively. In doing this, we first initialize the system randomly and then run the dynamics, for a given $p$, as described above until the model reaches a steady state. Following this, we calculate the above mentioned quantities and measure their time averaged value. The angular bracket then denotes the average over the groups or constituencies of the different quantities. 

The probability distribution of the winning margins and the other quantities behave very differently, depending upon whether $p<p_{c_1}$ or $p>p_{c_1}$ i.e., the consensus and the fragmented state. Figs. \ref{fig2} and \ref{fig3} shows the distributions of $M$ and $\mu$ for the mean-field and two-dimensional versions of the model for $p$ values on either side of the consensus to fragmented transition. After the transition point, where the opinion values are in a fragmented state, the distributions match what was reported in most of the election data in Ref. \cite{san}. However, the results for $p<p_{c_1}$ look similar to the rather non-transparent elections results, as discussed there. For higher values of $p$, the distributions do not significantly change further. Of course, it is worth recalling that for the mean-field version there is no further fragmented to polarized transition, which can be seen for the square lattices. 

In this context, it is also worth discussing the turnout fraction. As can be seen from Fig. \ref{fig4},the fraction of the neutral agents show a symmetric non-monotonic behavior about $p=1/2$ for the square lattice, reaching a maximum value of $1/3$ at $p=1/2$, when expectedly the opinion values are fully random. For the mean-field version, as was noted in Ref. \cite{san}, the only valid solution of $f_0$ for $p=p_{c_1}$ (or simply $p_c$, as there is no second transition here), is $1/3$. The simulation results verify that. In all cases, by definition, the turnout fraction is simply $1-f_0$. The distributions of the turnout fractions are Gaussian, with slight asymmetry for higher values of $p$, but is not of particular interest here, since we have simply kept all constituencies to be of equal sizes.  

The most prominent feature of the results here is that when there is a consensus formed in the system, the winning margin distribution looks qualitatively different from the case when there is no such consensus.

\section{Comparisons with election data in the US and the UK}
Of course, following the observations from the model simulations discussed so far, it is then important to compare with real examples of election data. As mentioned above, there is a qualitative difference in the shape of the winning margin distribution between the cases of consensus and fragmentation of opinion values. In order to look into that difference, we consider the elections results of five most prominent (above 10\% victory margins) election victories in the U.S. in the past 100 years (1936, 1956, 1964,1972 and 1984) and five of the most tightly contested (below 3\% victory margins) elections (1960, 1968, 1976, 2000 and 2004). Similarly, we look at the data from the three most prominent victories in the U.K. elections (1935, 1983 and 1997) and two of the most contested elections (1951 and February 1974). As mentioned before, the turnout is considered as the voting fractions of the two major parties -- Democrats and Republicans in the U.S. and Labour and Conservative in the U. K. Other voters are also considered as neutral, along with the registered voters who did not vote. We calculate the margins from the combined data of each group. For the U.S. elections, the data are at the state level (can be found in Ref. \cite{us_source}) and for the U.K. elections, the data are at the constituency level (can be found in Ref. \cite{uk_source}). From Fig. \ref{fig5}, it can be seen that for the U.S. data, a clear difference in the distribution functions appear between the land-slide victories and tightly contested elections, representing the emergent consensus ($p<p_{c_1}$) and fragmented state ($p>p_{c_1}$) in the model. For the U.K., the difference is not that prominent. One reason could be that in the first-past-the-post system in the U.K., there are significant voting fractions going to the other candidates than those from the two major parties, and therefore it does not clearly represent the consensus formation pictured in the model. For this reason, other examples of the U.K. elections could not be considered, and same for the Indian elections. This could perhaps require a modified version of the model, where the choices are not just binary. In the present context, even such modifications are not straightforward as any such third party consideration would require determining their position in the 'political spectrum' somewhere other than the two extreme ends. Such provisions are beyond the scope of the present work.

However, it is worth emphasizing that when a consensus is actually formed (demonstrated through land-slide victories), the nature of the margin distribution changes, in the same way it does in the model. 

\section{Discussions ans conclusions}
Efforts to model emergent consensus or lack thereof in societies has been a long standing effort. The developments of many models have shown that through rather simple interaction rules, emergent consensus could be seen. However, it is also important to make quantitative comparisons of the realistic features in those models with data from elections or similar such exercise, for example opinion surveys. 

The kinetic exchange opinion model is one such variant where emergent consensus could easily be modeled with simple interaction rules, similar to the wealth exchange models. It also shows several realistic features that were noted in real election data. 

Here we have shown that in terms of the distribution of margins, the model shows two distinct types of distributions, seen on either side of the consensus to fragmented state of opinions. Similar feature has recently been noted in election results of many countries. However, we note here that the departure from the universal feature noted in Ref. \cite{san} is not necessarily due to fraudulent practices. In general, it could happen whenever there is a very prominent consensus developed in the voting population. Of course, such consensus might be genuine opinion, as shown specifically for some of the U.S. elections, or it could be distortions, as noted in the examples cited in Ref. \cite{san}. In the former case, it would be useful to look into more fine-grained data and possibly for similar other examples elsewhere. It would then be useful to make a more direct connection with the model parameter with the features observed in real data. 

In conclusion, the kinetic exchange opinion model shows a rich behavior in terms of the increased noise parameter; starting with the much studied consensus to fragmented opinion transition and then again a transition to completely polarized state with anti-ferromagnetic like order on a square lattice, within the Ising universality class. The model also qualitatively reproduces properties of the winning margin distribution in various constituencies in the cases of land-slide victories (consensus state), as well as for tightly contested elections, demonstrated through data analysis of the U.S. and the U.K. elections.  

\section*{Acknowledgement}
SB acknowledges discussions and comments on the manuscript by Parongama Sen. Some of the computations were performed in HPCC Chandrama at SRM-AP.

\bibliographystyle{ws-ijmpc}
 
\end{document}